# Status of ITER Neutral Beam Cell Remote Handling System

N Sykes[a], C Belcher [d], C-H Choi[b], O Crofts[a], R Crowe[d], C Damiani[c], S Delavalle[d], L Meredith[d], T Mindham[a], J Raimbach[a], A Tesini[b], M Van Uffelen [c]

[a]CCFE. Culham Science Centre, Abingdon, OX14 3DB, UK,
[b]ITER Organisation, CS90 046, 13067 St. Paul les Durance Cedex, France,
[c]Fusion for Energy, C/Josep Pla 2, Torres Diagonal Litoral-B3, E-08019 Barcelona – Spain,
[d] Oxford Technologies Ltd, Abingdon, OX14 1RJ

The ITER neutral beam cell will contain up to three heating neutral beams and one diagnostic neutral beam, and four upper ports.

Though manual maintenance work is envisaged within the cell, when containment is breached, or the radiological protection is removed the maintenance must be conducted remotely. This maintenance constitutes the removal and replacement of line replaceable units, and their transport to and from a cask docked to the cell. A design of the remote handling system has been prepared to concept level which this paper describes including the development of a beam line transporter, beam source remote handling equipment, upper port remote handling equipment and equipment for the maintenance of the neutral shield. This equipment has been developed complete the planned maintenance tasks for the components of the neutral beam cell and to have inherent flexibility to enable as yet unforeseen tasks and recovery operations to be performed.

Keywords: Remote Handling Manipulator transporter Neutral Beam

## 1. Introduction

The reference designs for this conceptual design work are elaborated by Choi[1], where four systems, the Monorail Crane, Beam Source Remote Handling Equipment (BS-RHE), Beam Line Transporter (BLT), Upper Port Remote Handling Equipment (UP-RHE) and Tools, and their required tasks are described. Additional requirements have been identified for remote opening and closure of the beam line vessel (BLV), maintenance of the neutron shield (NS). The articulation of all these systems, concepts for the provision enabling services, tooling designs that they deploy and their remote recovery strategies are elaborated.

The maintenance of the NS is particularly challenging, given the radioactivity, its mass, and its inaccessibility within the vacuum vessel port extension. The solution developed for this involves a series of tools and equipment to be deployed from within the heating neutral beam line vessels, via a series of self installed rails to unfasten it, transport it into casks, while containing contamination.

The design of the overhead monorail crane is ostensibly described in O Crofts [2] and therefore is not repeated here.

## 2. Beam Line Transporter

### 2.1 Articulation

The conceptual design of the BLT[3] consists of powered swung and fixed rails mounted to the building, on which boom carriage is attached. This carriage provides the mating component for the remotely deployed boom and powers the traverse along the rails.

The articulated boom comprises of three link sections with actuated joints. The far end of the boom contains a pitch and roll joint onto which the mast hinge assembly is attached.

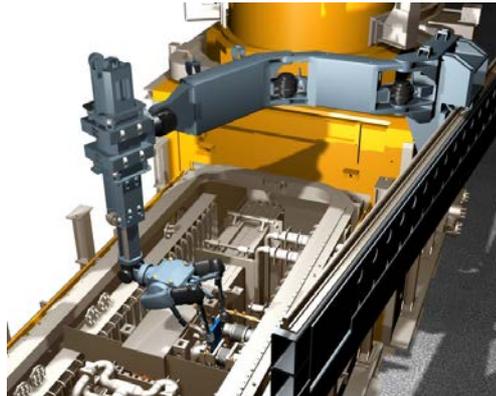

Fig 1 Beam Line Transporter maintaining RID

A mast assembly remains attached to this roll joint. A rack and pinion within it is used to drive the outer telescopic joint. The inner joint is directly driven by the movement of the outer with dual cables connected to the hinge assembly, running around rollers on the upper mast and finally connected to the lower mast. A driven gimbal joint is attached to the end of the mast and provides yaw, pitch and roll maneuverability to the manipulator or other end effectors. It consists of three driven axes with 180° of rotation on each joint.

The manipulator is a dexterous servo-mechanism, part of a master-slave, man in the loop system enabling mechanical tasks to be carried out remotely. The manipulator has two each of which has seven degrees of freedom.

A remotely powered winch with a lift capacity of 100kg can be installed on the front plate of the manipulator and is interchangeable with an articulated arm with a capacity of 50kg, for support of components in an overhead position.

## 2.2 Services

The service feeds supply all of the electrical components in the BLT system with the required power and data supply. Cable chain is coiled up with the aid of roller drums. The drums are free to turn and are driven on a lead screw to minimize the strain on the boom carriage.

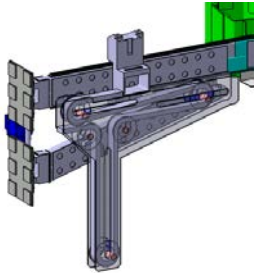

Fig 2 Services Management System.

## 2.3 Recovery

The failure modes detailed below are considered to be the most complex to recover [4].

Failure of the carriage drive can be recovered via a remote release and engagement of redundant system.

Failure of the articulating joints are recovered through a built in remote release of the joint, and then using the attached manipulator to assist withdrawal from area, before orientating the failed joint for crane recovery using surrounding structure.

In the case of a failure of the pitch or roll joints, the latching mechanism between the roll joint and the mast can be remotely disconnected.

Failure of the mast extending joint is resolved by remotely slackening the cable on the mechanism allowing it to slip then recover utilizing the remaining one.

In the event of a failure of the service management system, the carriage is moved back, so that excess cable chain will loop on the floor and be manually repaired after containment is restored.

## 2.4 Tooling

Requirements for the following tooling, to be operated by the manipulator (mounted on any equipment), have been identified as below.

**Table 1 Tooling Requirements for Neutral Beam Cell**

| Tooling | Description |
|---|---|
| Pipe tooling | |
| Bellows compression tool | 3 suites of tools – DN200, DN80/100 and DN25/50. Pipe inspection tools may, if required, include volumetric inspection tools |
| Pipe alignment tool | |
| Pipe cutting tool | |
| Pipe surface finish tool | |
| Pipe welding tool | |
| Pipe inspection tools | |
| Pipe leak test tool | |
| Flange tooling | |
| Flange carriage | Two varieties; manipulator operated and fully autonomous. |
| Lip seal cutting tool | |
| Lip seal welding tool | |
| Lip seal inspection tool | |
| Lip seal leak test tool | |
| Flange bolt runner tool | Fully autonomous only |
| Flange bolt torque tool | |
| Bolt tooling | |
| Torque multiplying bolt runner | Interchangeable sockets |
| Power torque tool | Bespoke tools for sizes M12..M36. |
| Layshaft | Extend manipulator reach |
| Miscellaneous tooling | |
| Radiation Tolerant Cameras | Configurations are dependent on specific tasks. |
| Conduit handling tool | Generic handling tool. |
| Vacuum Cleaner | |
| Task Module | For transportation of tools and services |
| Control services | |

## 3. Beam Source Remote Handling Equipment

### 3.1 Articulation

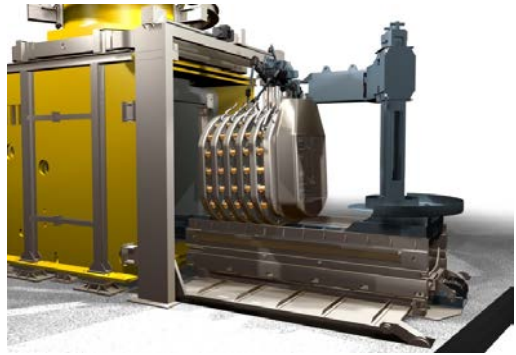

Fig 3 Beam Source Remote Handling Equipment extracting the source from the beam source vessel.

The conceptual design of the BS-RHE[3] comprises a support frame, a manipulator carriage, mast, articulated boom, and manipulator to provide dexterous manipulation and extending rails and source carriage, to enable the extraction and installation of the source.

The support frame incorporates slide-ways and racks for both the extending rails and the manipulator carriage, as well as a rear plug connection point.

The manipulator carriage contains the mast turntable and the pinion drive to traverse the support frame. The mast mounted on the turntable incorporates a rack on which the boom carriage is driven. The boom links, and manipulator are similar to those described in section 2.1.

The extending rails similarly contain a pinion drive to enable them to extend and dock with rail supports installed with the beam source vessel (BSV) and contain the slideways and rack for the source carriage.

The source carriage includes a pinion drive to traverse the extended rails and a hydraulically actuated source interface plate to lift and lower the beam source from its installed position.

## 3.2 Services

Rear plugs are required to be permanently installed behind each beam line, to provide automatic power and signal connection to the BS-RHE Frame.

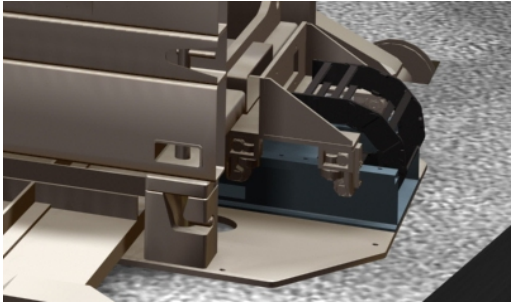

Fig 4 Rear Plug Connection

## 3.3 Recovery

The significant recovery scenarios are those where recovery action is required before the system can be removed from behind the beam line so that the rear passive magnetic shield can be closed. Three scenarios have been considered; failure of one of the articulated boom joints, failure of one of the drives in the mast carriage and failure of the rear plug mechanism.

The boom joint failure is recovered by the remote disconnection of the drive and utilizing the vessel structure to guide the failed joint into a suitable position for recovery.

The mast carriage drives are accessible and can be replaced with new by the manipulator.

The failure of the rear plug mechanism to disengage relies on a secondary drive to force the plug and socket apart.

## 3.4 Tooling

The manipulator can utilize the tooling described in section 2.4. In addition, the following tooling is used in the Cs Oven maintenance.

Table 2 Tooling Requirements for Cs Oven maintenance

| Tooling | Description |
|---|---|
| Gimbal Tube | Aligns tooling to Cs oven |
| Flight Tube | Supports tooling tube |
| Gripper Block | Provides winch connection |
| Tooling Tube | Disconnects and extracts the Cs Oven |
| Stillage | Enables transportation on the crane |

## 4 Upper Port remote Handling Equipment

### 4.1 Articulation

The conceptual design of the UPRHE[3] consists of following subassemblies, a toroidal transporter which provides the means of translating around the balcony rails and contains combined bearings and a drive for the radial extension trolley. The radial extension trolley is driven on a linear rack up to the upper port.

Mounted on the radial trolley is a tilt ramp which aligns the cask with upper ports and elevates to 11° incline using linear jacks. This tilt ramp bears diagnostic tube and a port cap containment casks, or a manipulator similar to that described in 3.1.

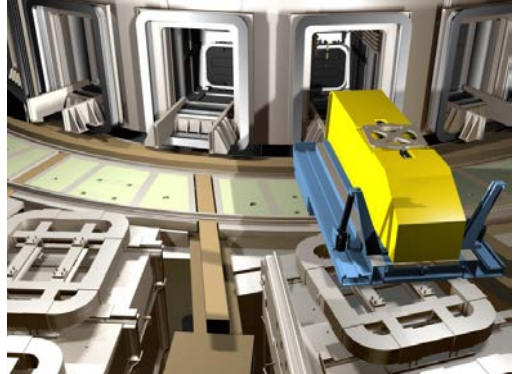

Fig 5 UP-RHE maintaining diagnostic tube

### 4.2 Services

A motorized cable management system releases the cable which is routed on top of the balcony plate. Mobile drums are moved along a lead screw actuated by a motor which is synchronized to the position of the toroidal trolley

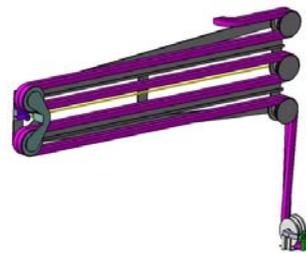

Fig 6 Upper Port Plug Cable management system

### 4.3 Recovery

The recovery of the toroidal trolley is enabled by a redundant motor. In case of failure of the extension rail motor, the recovery is ensured by a gearchain, which drives the extension rail and is accessed at the back by the closest BLT. In case of a tilting linear actuator failure, the recovery consists of driving the radial extension trolley back synchronized with the working remaining jack. The actuators are not back drivable so the weight of the tilt ramp will push the faulty jack against the end stop and restore it to a horizontal recoverable position.

## 5 Vessel Opening Systems

### 5.1 Articulation

The lid lifting device lowers and locks to the BLV lid. Linear actuators lift the lid vertically to clear the alignment features. The cable drive starts to retract the cable lifting the lid, then linear actuators close the parallel mechanism to reduce the rotation radius and compress the stowed volume.

The rear flange equipment turns about the pillar and attaches to the BSV rear flange. Linear actuators move the flange horizontally to clear alignment features, before the whole assembly rotates 90° about the support pillar and the parallel mechanism is closed to allow access into the BSV.

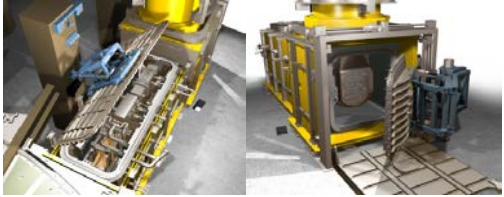

Fig 7: Vessel Lid Remote Handling Equipment & Vessel Rear Flange Remote Handling Equipment

### 5.2 Services

These systems are permanently hard wired .

### 5.3 Recovery

The equipment has extensive redundancy in the design. The areas not covered by this are lid lifter recovery procedures concerning the latch mechanisms and the vertical drives which are exchanged by the BLT.
The door opener has three identified recovery procedures for hinge rotational, door latching, intermediate locking failures where minor design alterations will allow exchange by BLT or BS-RHE.

## 6. Neutron Shield Maintenance System

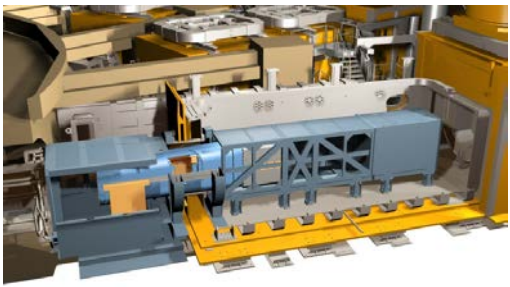

Fig 8 NS Remote Handling System

### 6.1 Articulation

A number of equipment items have been conceptually designed to fulfill this operation. A series of contamination confinement casks fitted with racked rails casks allow the storage and transportation of the shield bearer, extension panel transporter, the extension panels (the NS panels which are upstream of the attachment flange in the port extension) and the NS itself. These rails allow translation from the casks installed on stillage within an emptied BLV, to the connecting tube. The connecting tube has a number of functions. It has vertical actuation to allow the transported items to be lifted and lowered between the level of the cask rails and those within the connecting duct, where the NS is installed. It installs the rails, using an indexing translation mechanism into both the BLV extension and the connecting duct. A preloaded bolting tool fastens these rails onto their respective rails supports. The extension panel transporter, docks with tooling to interface with top, bottom, and side extension panels. It has incorporated rotation to allowing indexing between these positions and liner actuation to extract the panels away from their installed position. The interface tooling is loaded with pre-positioned drives to fasten the tool to the panels and to unfasten them from their supports. The transporter is also fitted with a drive to translate along the installed rails. The shield bearer is self propelled with 5 driven axes giving 6 degrees of freedom for maneuvering the NS. It has a bolting system to unfasten the securing bolts, and can be adapted to support pipe welding systems.

### 6.2 Services

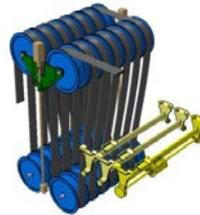

Fig 9 NS Cable Management System

A motorized cable reeler is installed in the casks, consisting of two rows of seven drums around which the length of cables are routed. Two motorized lead screws move the top drums up and down to reel the cable.

### 6.3 Recovery

Due to the radiation levels and containment requirement for handling the NS, if a breakdown occurs within the containment path, neither manned access nor support of BLT will be possible, therefore the vertical linear actuation of the connecting tube has redundancy. The shield bearer will be equipped with dual drives on all six axes, each with capacity to power individually. A large overcapacity in the bearing specification will reduce the risk of bearing failure. A secondary feed to the critical drives is required. Similarly both axis requiring recovery of the extension panel transporter have a redundancy and if the main rotation drive seizes a remote disconnection will allow descent to a recoverable position.

## 7. Conclusions

Feasible designs of equipment have been established for the currently foreseen maintenance tasks within the neutral beam cell. Recovery procedures have been devised for single point failures of this equipment

## 8 Acknowledgments

This work was funded jointly by the RCUK Energy Programme under grant EP/I501045 and by F4E under grant 51. The views and opinions expressed herein do not necessarily reflect those of F4E or European Commission or IO. F4E is not liable for the use which might be made of the information in this publication.